\documentclass[pra,aps,floats,letterpaper,floatfix,footinbib,preprintnumbers]{revtex4}
\usepackage{amsmath}
\usepackage{hyperref}
\usepackage{color}
\usepackage[mathenv]{}
\usepackage{graphicx}
\usepackage{ulem} 
\usepackage{natbib}
\begin{document}

\title{Amplified Dispersive Optical Tomography}

\author{Keisuke Goda}
\author{Daniel R. Solli}
\author{Bahram Jalali}

\affiliation{Department of Electrical Engineering, University of California, Los Angeles, California 90095, USA}


\begin{abstract}
Optical coherence tomography (OCT) \cite{huang1991science,fujimoto2003naturebiotech} has proven to be a powerful technique for studying tissue morphology in ophthalmology \cite{fercher1995optcommun,swanson1993optlett}, cardiology \cite{fujimoto2003naturebiotech}, gastroenterology \cite{tearney1997amjgastroenterol,yun2006naturemed,fujimoto2003naturebiotech}, and endomicroscopy \cite{adler2007naturephoton}. Its performance is limited by the fundamental trade-off between the imaging sensitivity and acquisition speed \cite{choma2003optexpress,fercher2003repprogphys,leitgeb2003optexpress} -- a predicament common in virtually all imaging systems. In this paper, we circumvent this limit by using distributed Raman post-amplification of the reflection from the sample. We combine the amplification with simultaneously performed dispersive Fourier transformation, a process that maps the optical spectrum into an easily measured time-domain waveform. The Raman amplification enables measurement of weak signals which are otherwise buried in noise. It extends the depth range without sacrificing the acquisition speed or causing damage to the sample. As proof of concept, \textit{single-shot} imaging with 15 dB improvement in sensitivity at an axial scan rate of 36.6 MHz is demonstrated. 
\end{abstract}

\maketitle


\section{Introduction}
Since the invention of OCT in 1991 \cite{huang1991science}, it has evolved into several different varieties, all of which can be placed into two major categories: time-domain OCT \cite{huang1991science} and frequency-domain OCT \cite{choma2003optexpress,yun2003optexpress,chinn1997optlett,adler2007naturephoton,wojtkowski2004optexpress}. Frequency-domain OCT provides higher acquisition speed and better signal-to-noise ratio (SNR) than the time-domain approach \cite{leitgeb2003optexpress,choma2003optexpress} because it avoids mechanical scanning. The frequency-domain OCT uses a broadband optical source, and the depth profile is encoded into the optical spectrum through the interference between the two arms of a Michelson interferometer. There are two implementations of this, which use different optical sources. In one implementation (the so-called swept-source OCT \cite{yun2003optexpress,chinn1997optlett,adler2007naturephoton}), the optical frequency is encoded in time with a frequency-tunable source. In the other implementation (the so-called Fourier-domain OCT \cite{fercher1995optcommun,wojtkowski2004optexpress}), the source has a wide instantaneous bandwidth and the spectrum is obtained using a diffraction grating combined with a linear detector array. In both cases, due to the Fourier relation between the temporal autocorrelation and the spectral power density, the depth information is obtained after an inverse Fourier transform of acquired spectra. 

Among the two frequency domain approaches, Fourier-domain OCT achieves better axial resolution as it benefits from the availability of supercontinuum sources with large bandwidth (hundreds of nm). However, Fourier-domain OCT is bulky and expensive, and has low environmental tolerance because it requires a diffraction grating and array detector (e.g. a CCD or CMOS) \cite{fercher1995optcommun,wojtkowski2004optexpress,leitgeb2004optexpress}. Moreover, the data acquisition speed of a CCD is low (only up to 10 kHz), limiting the imaging speed. In other words, there is a trade-off between the spectral resolution and imaging speed. On the other hand, swept-source OCT does not require an optical spectrometer because the wavelength is known for each time sample; however, the bandwidth, and hence the axial resolution, is limited by the tuning range of the laser, and the broadening of the instantaneous linewidth limits the depth range.

Dispersive Fourier transformation (FT) exploits the mathematical equivalence between paraxial diffraction and temporal dispersion \cite{solli2008naturephoton}. Its ability to map the spectrum into a temporal waveform has been used in measurement of the spectrum of laser pulses \cite{fetterman1979applphyslett,tong1997electronlett}, fiber dispersion measurement \cite{tong1997electronlett,hult2007jlightwavetechnol}, absorption spectroscopy \cite{chou2004ieeephotontechnollett,hult2007optexpress,chou2007LEOS}, Raman spectroscopy \cite{solli2008naturephoton}, reflectometry \cite{saperstein2007optexpress}, and OCT \cite{moon2006optexpress}. Measuring the spectrum in Fourier-domain OCT using dispersive FT avoids these issues by eliminating the diffraction grating and detector array \cite{saperstein2007optexpress,moon2006optexpress}. These elements are replaced by a dispersive fiber, a single detector, and a digitizer. This simplifies the system, and more importantly, enables fast real-time image acquisition. However, the loss in the dispersive medium, which, at the most basic level, is caused by the intimate connection between dispersion and loss described by the Kramers-Kronig relations, \cite{jackson1999wiley} limits the imaging sensitivity as well as the depth range. The latter can be understood by recognizing that, by virtue of spectrum-to-time conversion, the spectral resolution (which determines the depth range) is fixed by the temporal resolution of the electrical detection system. Stated differently, the electrical bandwidth of the digitizer limits the spectral resolution, a relation given by $\Delta f = 0.35\lambda_0^2/cDf_{\rm dig}$, where $\Delta f$ is the spectral resolution, $\lambda_0$ is the center wavelength, $c$ is the speed of light in vacuum, $D$ is the total group-velocity dispersion, and $f_{\rm dig}$ is the input bandwidth of the real-time digitizer. The product of $\Delta f \times D$ is fixed by the bandwidth of the digitizer; hence, to increase the optical resolution (i.e., to increase the depth range) one is forced to increase the total group-velocity dispersion, $D$. But this comes at the expense of increased optical loss and reduced detection sensitivity. The loss in the dispersive element is therefore the central problem in high sensitivity detection.  This implies that a longer integration time must be used, limiting the axial scan rate. Therefore, the loss in the dispersive element creates a trade-off between the sensitivity and scan rate. Increasing the laser power is not an attractive solution because it can cause damage to the tissue \cite{konig1997optlett}.

Our approach is different from previously demonstrated dispersive Fourier-domain OCT through the use of internal amplification in the dispersive element. By compensating for the loss in the dispersive element, it overcomes the trade-off between the imaging sensitivity (and hence the imaging depth) and acquisition speed. For optimum performance, we perform this function in the sample arm of the Michelson interferometer (Fig. 1) in order to increase the strength of the signal reflected from the sample, which enhances the interference contrast in the interference fringe and improves the imaging sensitivity. Internal amplification in a dispersive element has previously been used to demonstrate real-time Raman \cite{solli2008naturephoton} and absorption spectroscopy \cite{chou2007LEOS}, and a femtosecond digitizer \cite{chou2007applphyslett}. 

The desirable features for a dispersive element are high total dispersion, low loss, large optical bandwidth, smooth dispersion over the bandwidth, and commercial availability. Dispersion compensation fiber (DCF) offers an optimum combination of these parameters and is our preferred choice. While the loss can also be compensated by discrete optical amplifiers (such as erbium-doped fiber amplifiers, or even semiconductor optical amplifiers), distributed Raman amplification within the dispersive DCF is superior because it maintains a relatively constant signal level throughout the FT process. This important property maximizes the signal-to-noise-and-distortion ratio by keeping the signal power away from low power (noisy) and high power (nonlinear) regimes. Incidentally, this advantage of distributed Raman amplification over discrete amplification is known in long haul fiber optic communication links \cite{islam2002ieeejseltopquantelectron}. Raman amplification has another significant advantage: in an amorphous medium such as glass, it is naturally broadband. The gain spectrum can be further tailored by using multi-wavelength pump lasers, and, surprisingly but fortuitously, extremely broadband gain spectra can be realized using incoherent pump sources \cite{solli2008naturephoton}. This is highly desirable because a large optical bandwidth results in high axial image resolution in OCT. Raman-amplified dispersive elements also eliminate the need for a high power source, which can potentially cause damage to the sample \cite{konig1997optlett} and unwanted nonlinear signal distortion \cite{bouma1996optlett}.

\begin{figure}
\includegraphics[angle=0, width=0.8\columnwidth]{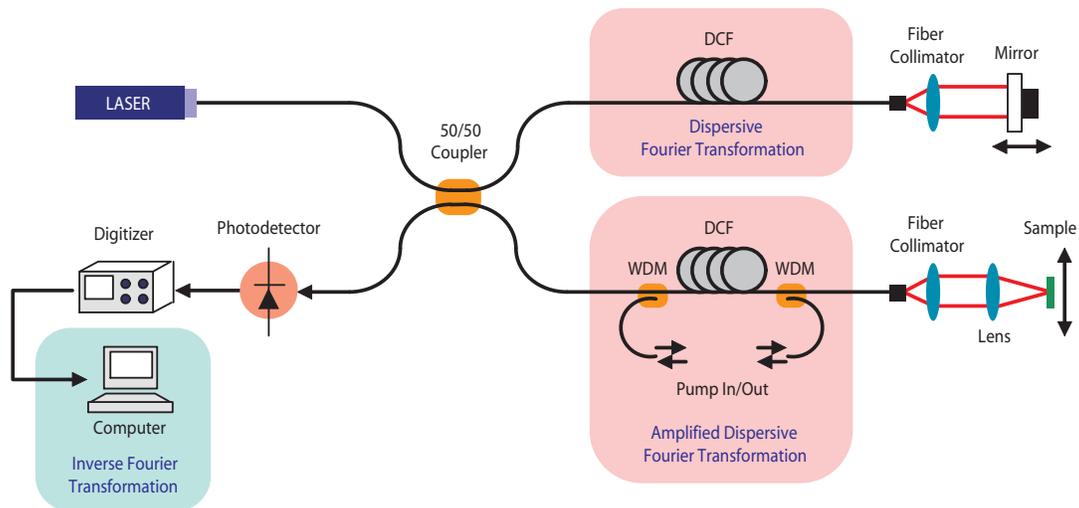}
\caption{Schematic of the experimental amplified dispersive optical tomography (ADOT) system. WDM: wavelength-division multiplexer; DCF: dispersion compensation fiber. The optical source is a mode-locked femtosecond laser with a repetition rate of 36.6 MHz. It is injected into the Michelson interferometer that contains a mirror and sample. The reflections from the mirror and sample interfere at the 50/50 fiber coupler. Each interferometer arm has a DCF module with equal round-trip dispersion of -1316 ps/nm and loss of 7.6 dB. Dispersive Fourier transformation is performed in the DCF of each arm, mapping the spectrum into a temporal waveform. The equal dispersion balances the dispersive Fourier transforms in the arms. During the dispersive Fourier transformation in the sample arm, distributed Raman amplification is implemented by pumping it with two diode lasers with center wavelengths of 1470 nm and 1480 nm. The Raman pumps are injected into and removed from the DCF by the WDMs. The output of the interferometer is detected by the AC-coupled photodetector with 50 ps response time, and captured with the 50 GS/s real-time digitizer with 16 GHz bandwidth. Inverse Fourier transformation is performed on the digitizer output to map the temporal waveform into the depth profile.} 
\end{figure}

\begin{figure}
\includegraphics[angle=0, width=0.7\columnwidth]{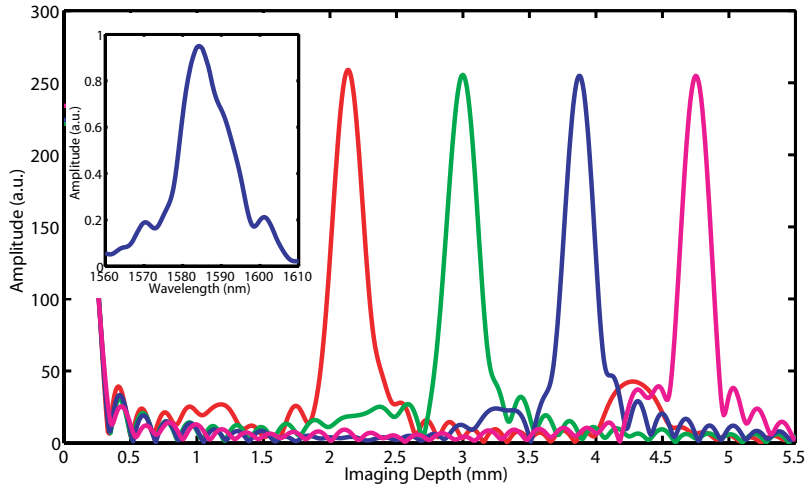}
\caption{Basic performance of the ADOT system showing the spectrum of the optical source and single-shot point spread functions at various imaging depths. After filtering and amplification, a spectrum centered at 1583.8 nm with a FWHM bandwidth of 15.0 nm is obtained. The axial resolution (227 $\mu$m) is limited by the modest bandwidth of the source (15.0 nm) centered at 1583.8 nm in this proof-of-principle demonstration, but this is not an inherent limitation of the technique, and can be significantly improved using a supercontinuum source which much broader bandwidth as long as temporal pulse overlap after dispersion does not occur.} 
\end{figure}

\section{Experiment}
The schematic of the amplified dispersive optical tomography (ADOT) is shown in Fig. 1. The optical source is a mode-locked femtosecond laser with 36.6 MHz repetition rate. Its spectrum is shown in Fig. 2. A pulse train with 2.7 kW peak power and 150 fs pulse width is injected into the Michelson interferometer where the reference and sample arms contain a mirror and sample at their ends, respectively. Each arm has a DCF module with equal round-trip dispersion of -1316 ps/nm and loss of 7.6 dB. The DCF performs dispersive FT to map the spectrum to a temporal waveform. The equal dispersion balances the dispersive FTs in the arms. The reflections from the sample and mirror interfere at the 50/50 fiber coupler, resulting in a spectral fringe in which the depth profile of the sample is encoded. Distributed Raman amplification is implemented in the DCF by pumping it with two 98.5 mW diode lasers with center wavelengths of 1470 nm and 1480 nm. The interferometer output is detected by an AC-coupled photodetector with 50 ps response time and captured with a 50 GS/s real-time digitizer with 16 GHz bandwidth. Inverse FT is performed on the digitizer output to map the time-domain waveform into the depth profile. The axial scan rate is equivalent to the repetition rate of the laser (36.6 MHz), which is the highest axial scan speed ever reported. It is two orders of magnitude faster than conventional OCT \cite{adler2007naturephoton}. 

The basic performance of the ADOT is shown in Fig. 2. \textit{Single-shot} ADOT point spread functions with a mirror in the sample arm at various imaging depths are evident in the figure. In this proof-of-principle demonstration, the axial resolution (227 $\mu$m) is limited by the modest bandwidth of the source (15 nm) centered at 1583.8 nm, but we emphasize that this is not an inherent limitation of the technique, and can be significantly improved using a supercontinuum source which much broader bandwidth. To avoid temporal pulse overlap -- after dispersion -- this will also require a proportional reduction in the pulse repetition rate. As an example, a source with 150 nm bandwidth will give an axial resolution of 22.7 $\mu$m (with a theoretical limit of 7.4 $\mu$m), while still achieving an axial scan rate of 3.66 MHz.

\begin{figure}
\includegraphics[angle=0, width=0.7\columnwidth]{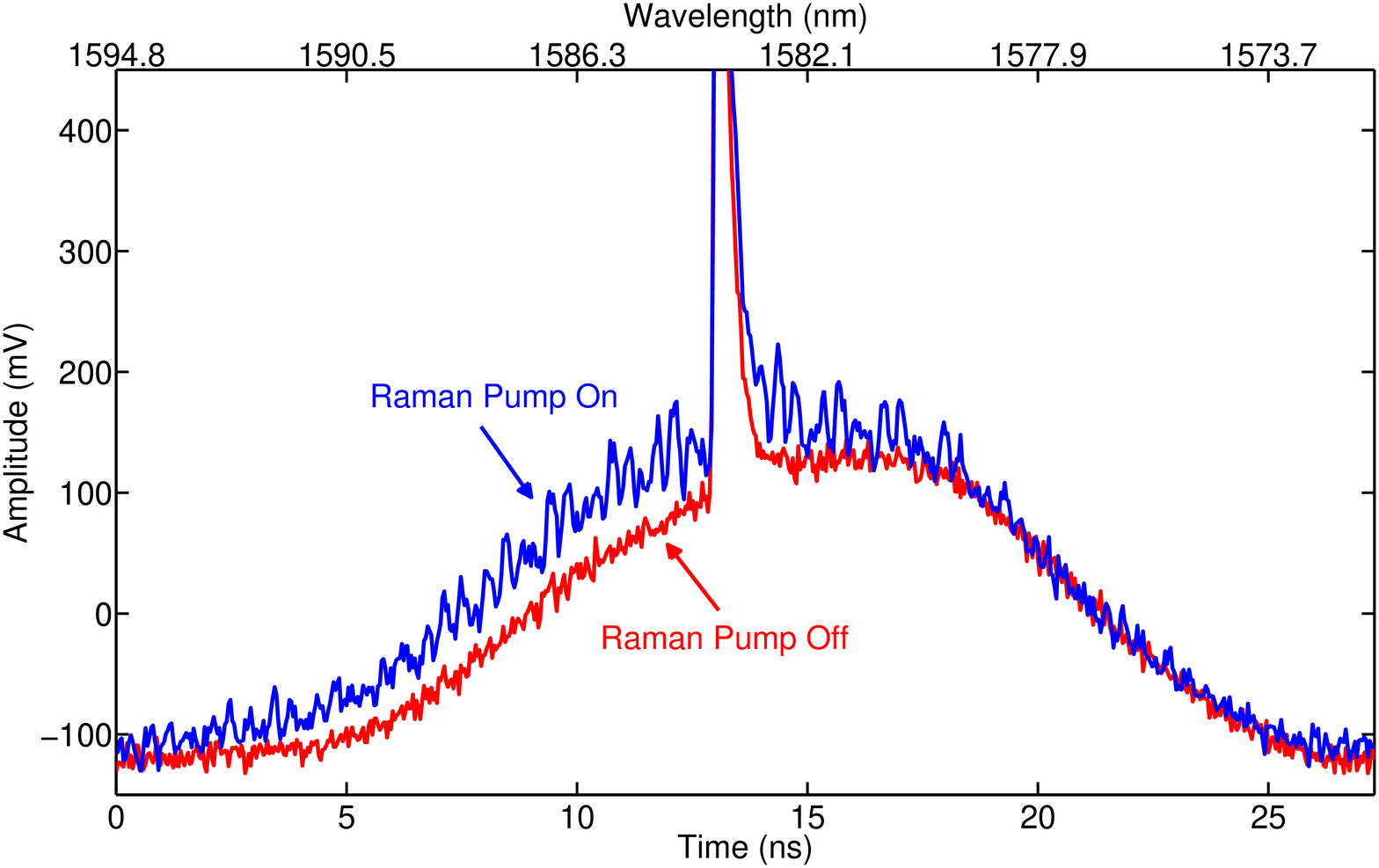}
\includegraphics[angle=0, width=0.7\columnwidth]{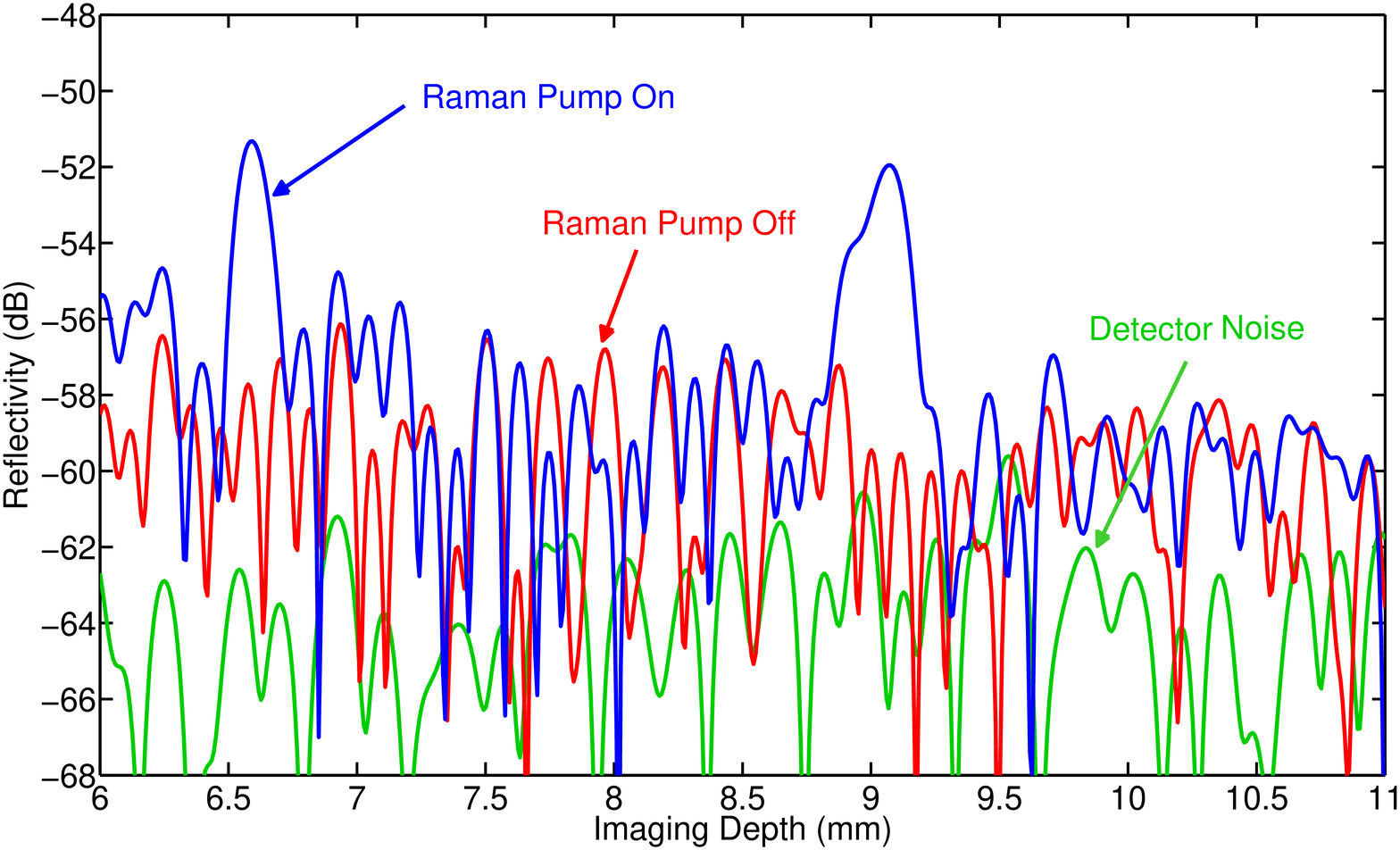}
\caption{Demonstration of the distributed Raman post-amplification to boost an, otherwise invisible, weak reflection signal. (a) Single-shot interference spectrum measured on a digital oscilloscope, with and without the Raman amplification. The sample is a partial reflector with two layers (both about -60 dB reflectivity) at imaging depths of 6.6 mm and 9.1 mm. The spectrum is mapped into the time-domain waveform by the dispersive Fourier transformation. Raman amplification improves the fringe visibility, which is otherwise invisible. The spikes at 13 ns are due to the reflection of the input beam from the 50/50 fiber coupler without going into the interferometer, and can be used as a trigger for the digitizer. The calibrated wavelength axis is shown above of the figure. The detector noise without any light incident on it is also shown in the figure. (b) Depth profile of the sample obtained by performing inverse Fourier transformation on the pulse with and without the Raman amplification. The depth profile only becomes visible with the amplification. Balanced detection \cite{adler2007naturephoton,choma2003optexpress}, which is used to cancel out common-mode noise by differential measurement of the two outputs of the 50/50 coupler, can also be incorporated into the ADOT system to further increase the sensitivity. The sensitivity can also be improved by multi-shot time-integrated detection and averaging.} 
\end{figure}

\begin{figure}
\includegraphics[angle=0, width=0.7\columnwidth]{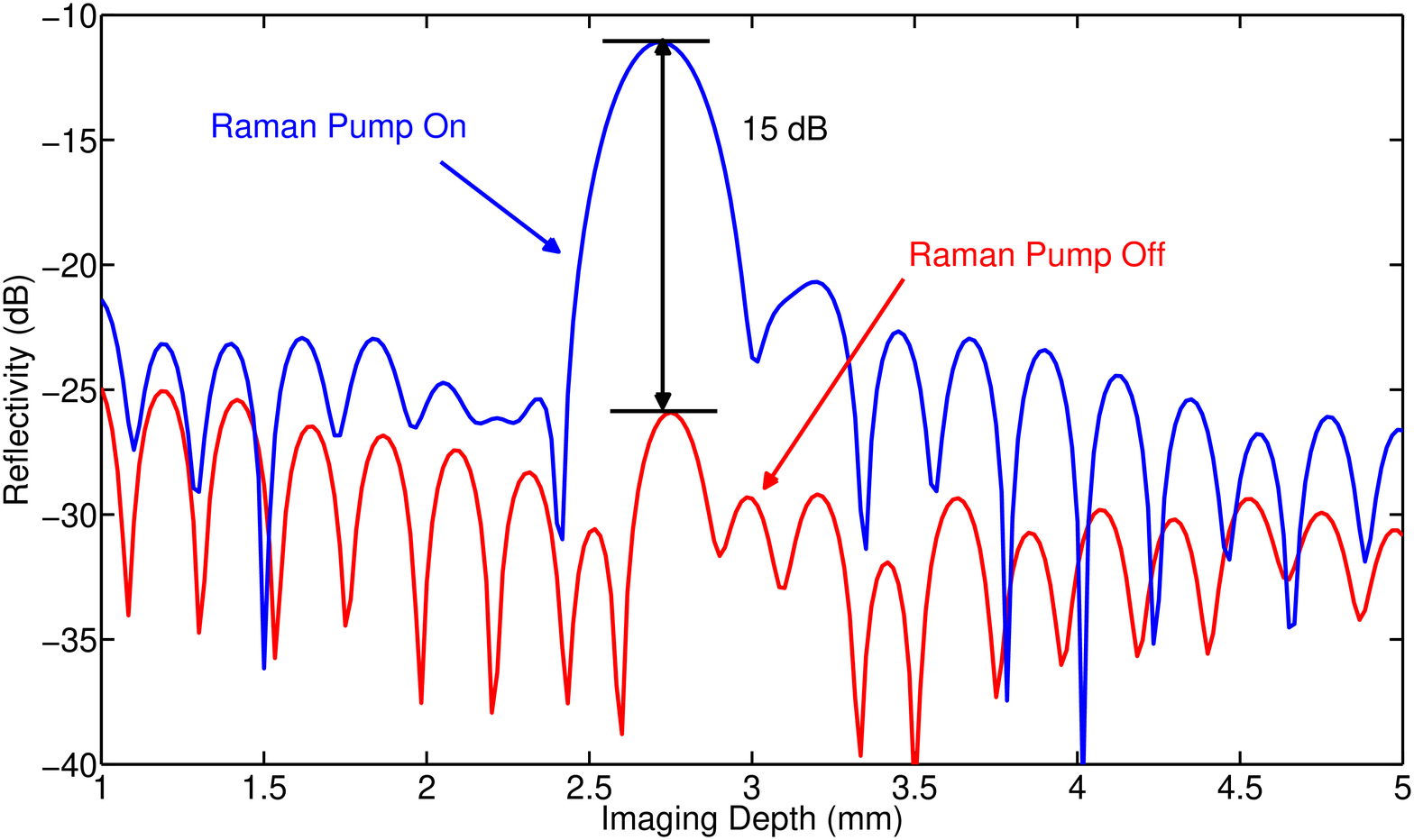}
\caption{Demonstration of an improvement in sensitivity due to the distributed Raman post-amplification. Single-shot depth profile of a -26 dB reflector at an imaging depth of 2.8 mm with and without the Raman amplification. An improvement in sensitivity by 15 dB is evident in the figure.} 
\end{figure}

Fig. 3a shows the \textit{single-shot} spectrum of an interferometer output pulse (with and without the Raman amplification) encoded using a sample with multiple reflecting layers. The amplification scheme utilizes two diode lasers to create a uniform Raman gain profile across the signal bandwidth. The amplification improves the fringe visibility of the interferometer, which is otherwise buried in noise. Fig. 3b shows the depth profile obtained by performing inverse FT on the pulse in Fig. 3a. The, otherwise invisible, depth profile becomes visible due to the amplification. Although signals in both directions (to and from the sample) pass through the amplifier, only the return signal is effectively amplified -- the incident signal is not amplified significantly because its peak power is much larger than the pump power supplied to the Raman amplifier. Fig. 4 shows the depth profile of a -26 dB reflector at an imaging depth of 2.8 mm with and without the Raman amplification. An improvement in sensitivity by 15 dB is evident in the figure, resulting in a SNR increase. Although the Raman gain bandwidth in this system is limited, broadband Raman amplification can be implemented for high-resolution OCT systems through the usage of additional Raman pump lasers or broadband incoherent pump light \cite{solli2008naturephoton}.


ADOT can be extended to any wavelength band where dispersive elements and Raman pump lasers are available. Balanced detection \cite{adler2007naturephoton,choma2003optexpress}, which is used to cancel out common-mode noise by differential measurement of the two outputs of the 50/50 coupler, can also be incorporated into ADOT to further increase the imaging sensitivity. The imaging sensitivity can also be improved by multi-shot time-integrated detection and averaging. Additionally, although the enhancement factor in sensitivity is limited to 15 dB in this proof-of-principle demonstration, the use of more powerful Raman pumps or a fiber with higher Raman gain coefficient can enhance the sensitivity significantly. To the best of our knowledge, this is the first time optical post-amplification has been used for improvement in imaging. 

\section{Acknowledgments}
This work was partially supported by DARPA. We are grateful to S. Gupta at UCLA for valuable discussions. 



\section{Appendix}
\subsection{Theory}
As described above, ADOT is based on amplified dispersive FT where the reflection signal from a sample is Raman-amplified during dispersive FT. The dispersive FT based on the second-order mode-propagation constant, $\beta_2$, has been characterized previously \cite{solli2008naturephoton}. For large bandwidths (over 100 nm), it is, however, important to include the higher-order mode-propagation constants. In this case, the mode evolution of a Fourier-domain signal $\tilde{U}(0,\omega)$ in an amplified dispersive element is given by
\begin{eqnarray}
\label{U}
U(z,T) = \frac{G}{2\pi}\int_{-\infty}^{\infty}\tilde{U}(0,\omega)\mbox{exp} \left[\frac{i}{2}\beta_2(\omega-\omega_0)^2z + \frac{i}{6}\beta_3(\omega-\omega_0)^3z-i(\omega-\omega_0)T\right]d\omega,
\end{eqnarray}
where $z$ is the dispersion length, $T$ is measured in a frame of reference moving with the pulse at the group-velocity $v_g = 1/\beta_1$ and given by $T=t-\beta_1 z$, $\omega_0$ is the center frequency of the optical source, $\omega$ is the measurement (sideband) frequency, $\beta_3$ is the third-order mode-propagation constant, and G is the gain factor (in amplitude) which is assumed to be constant. If the group-velocity dispersion is large, applying the stationary-phase approximation \cite{solli2008naturephoton} to Eq. \ref{U} yields
\begin{eqnarray}
\label{absU}
|U(z,T)| \simeq\frac{G}{2\pi}|\tilde{U}(0,\omega)|,
\end{eqnarray}
where 
\begin{eqnarray}
\label{omega_T}
\omega = \omega_0 -\frac{\beta_2}{\beta_3}+\sqrt{\left(\frac{\beta_2}{\beta_3}\right)^2+\frac{2T}{\beta_3 z}}\simeq \omega_0 + \frac{T}{\beta_2 z}-\frac{\beta_3T^2}{\beta_2^3z^2}.
\end{eqnarray}
Eqs. \ref{absU} and \ref{omega_T} indicate the transformation of the input spectrum into an amplified temporal waveform. The approximation holds as $2\beta_3T/\beta_2^2z\ll 1$, and agrees with the dispersive Fourier transformation with only the second-order mode-propagation constant considered \cite{solli2008naturephoton}. The transformation can also be expressed in terms of the dispersion parameter, $D$, and its dispersion slope, $dD/d\lambda$, 
\begin{eqnarray}
\Delta T(\lambda)= D(\lambda_0)z\Delta \lambda + \frac{1}{2}\left.\frac{dD}{d\lambda}\right|_{\lambda_0}z(\Delta \lambda)^2,
\end{eqnarray} 			       
where $\Delta \lambda=\lambda-\lambda_0$ is the bandwidth of the optical source, $\lambda_0 = 2\pi c/\omega_0$ is the center wavelength, $\lambda = 2\pi c/\omega$ is the measurement wavelength, $D(\lambda_0)=-2\pi c\beta_2/\lambda_0^2$ is the dispersion parameter evaluated at $\lambda=\lambda_0$, and $\left.dD/d\lambda\right|_{\lambda_0}=4\pi c \beta_2/\lambda_0^3+(2\pi c)^2 \beta_3/\lambda_0^4$ is the dispersion slope evaluated at $\lambda=\lambda_0$. In this experiment, $D(\lambda_0)z=-1316$ ps/nm, $\left.dD/d\lambda\right|_{\lambda_0}z=-0.46$ ps/nm$^2$, and $\Delta\lambda=15.0$ nm, and therefore, the dispersion slope is of little importance. However, for optical sources with large bandwidths, the effect of the dispersion slope on the amplified dispersive FT needs to be taken into account. 

\subsection{Experimental Details}
In our experiment, the light from the femtosecond mode-locked laser (Precision Photonics) is filtered and amplified by an erbium-doped fiber amplifier (PriTel), and a broadband spectrum with a center wavelength of 1583.8 nm and a FWHM bandwidth of 15.0 nm is obtained. A variable attenuator (Thorlabs) and a polarization controller (Precision Photonics) are used in the reference arm to optimize the fringe formed by the reflections from the sample and reference arms. The mirror in the reference arm is placed on a translation stage with a micrometer actuator (Newport) to adjust time delay between the return pulses in the sample and reference arms. The sample in Fig. 3 consists of a weakly reflecting transparent thin film and a weakly reflecting mirror which is located 2.5 mm apart from the film. The incident light is focused onto the film so that the mirror reflection coupled back into the fiber is about -60 dB of the light incident on it. The sample in Fig. 4 is also a weakly reflecting mirror. The sample is mounted on a translation stage in the transverse direction. The Raman pumps used to pump the DCF module are diode lasers (Furukawa) designed for distributed Raman amplification in telecommunications systems. The photodetector is an AD-50ir amplified photodetector (Newport) with a noise-equivalent-power of 15 pW/$\sqrt{\rm Hz}$. The digitizer is a DPO71604 oscilloscope (Tektronix). 

Based on the group-velocity dispersion of -1316 ps/nm and the digitizer sampling rate of 50 GS/s, the spectral resolution of the ADOT system is found to be $\delta \lambda$ = 30.4 pm (assuming that at least two sampling points are required to resolve spectra), which corresponds to a depth range of 20.6 mm in air. 




\end{document}